\documentclass[superscriptaddress,aps,prl,twocolumn,showpacs,amsmath,amssymb]{revtex4-1}
\usepackage{graphicx}
\usepackage{color}

\begin{document}
\title{Strongly anisotropic ballistic magnetoresistance in compact three-dimensional semiconducting nanoarchitectures}

\author{Ching-Hao Chang}
\email{cutygo@gmail.com}
\affiliation{Institute for Theoretical Solid State Physics, IFW Dresden, Helmholtzstr. 20, 01069 Dresden, Germany}
\affiliation{Department of Physics, National Tsing Hua University, Hsinchu 30043, Taiwan}
\author{Jeroen van den Brink}
\affiliation{Institute for Theoretical Solid State Physics, IFW Dresden, Helmholtzstr. 20, 01069 Dresden, Germany}
\affiliation{Department of Physics, Technical University Dresden, D-1062 Dresden, Germany}
\author{Carmine  Ortix}
\affiliation{Institute for Theoretical Solid State Physics, IFW Dresden, Helmholtzstr. 20, 01069 Dresden, Germany}


\begin{abstract}
We establish theoretically that in non-magnetic semiconducting bilayer or multilayer thin film systems rolled-up into compact quasi-one-dimensional nanoarchitectures, the ballistic magnetoresistance is very anisotropic: conductances depend strongly on the direction of an externally applied magnetic field. This phenomenon originates from the curved open geometry of rolled-up nanotubes, which leads to a tunability of the number of one-dimensional magnetic subbands crossing the Fermi energy. The experimental significance of this phenomenon is illustrated by a sizable anisotropy that scales with the inverse of the winding number
, and persists up to a critical temperature that can be strongly enhanced by increasing the strength of the external magnetic field or the characteristic radius of curvature, and can reach room temperature. 
\end{abstract} 

\maketitle

In 1857 Thomson discovered that the resistivity of bulk ferromagnetic metals depends on the relative angle between the electric current and the magnetization direction \cite{Thomson}. The prediction of this anisotropic magnetoresistance (AMR), caused by an anisotropy in the electron scattering due to spin-orbit interaction, was experimentally verified more than a century later in iron, cobalt, and nickel  alloys. Since then, the interest in this phenomenon has received a boost thanks to the development of AMR sensors for magnetic recording \cite{amr-fe, amr-book}. 

It was recently proposed that this phenomenon might also occur in ferromagnetic Fe and Ni nanowires \cite{bal-ma}.
Contrary to macroscopic samples, in miniaturized objects with characteristic dimensions less than the electronic mean free path, electronic transport is ballistic rather than diffusive, and electron scattering does not contribute to the conductance. 
The ballistic conductance of a quasi one-dimensional (1D) magnetic nanostructure with a slowly varying constriction of width of the order of the Fermi wavelength $\lambda_F$ is indeed simply given by $G= N e^2 / h$ where $N$ is the number of open conducting channels \cite{qnc-q}. Due to the strong spin-orbit coupling in these nanostructures, the number of transverse modes at the Fermi energy changes with the magnetization direction and leads ultimately to a ballistic anisotropic magnetoresistance (BAMR).

In this Letter, we show that a strong BAMR occurs in the electronic quantum  transport of  compact quasi-1D  tubular nanostructures obtained by the self-rolling of strained thin films \cite{firstn}, when subject to an externally applied magnetic field. Contrary to the nanostructures of magnetic materials mentioned above, the BAMR in these nanoarchitectures is entirely due to the {\it open curved} geometry of the rolled-up tubes, which  breaks the rotational symmetry of the three-dimensional (3D) embedding space. As a result, rolled-up tubes of conventional {\it non-magnetic} semiconducting materials 
display a sizable BAMR, which scales with the inverse of the number of windings of the tubes, and persists up to a temperature that can be tuned to be as large as room temperature by tailoring either the strength of the magnetic field or the radius of curvature.  

\begin{figure}[tb]
\includegraphics[width=.9\columnwidth]{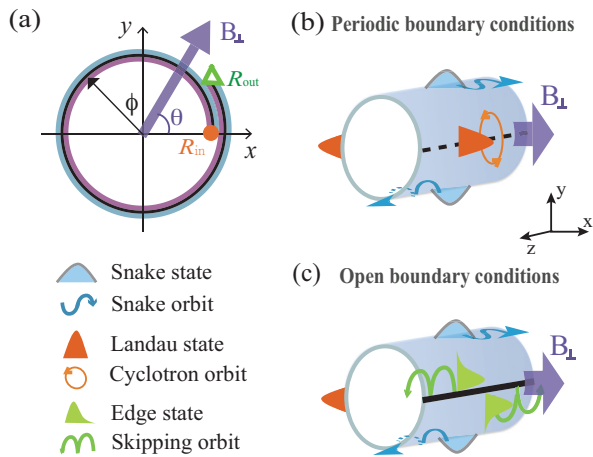}
\caption{(Color online). (a) Cross section of a rolled-up nanotube with 1.1 windings perpendicular to the tube axis. $R_{in}$, $R_{out}$ indicate the inner and outer radii of curvature. The black line indicates the 2DEG embedded in a bilayer stack. (b) Cartoon representation of the magnetic states for a 2DEG on a nanotube with periodic boundary conditions. (c) Same for open boundary conditions.}
\label{fig:fig1}
\end{figure}

Our starting point is 
a curved two-dimensional electron gas (2DEG) in a GaAs heterojunction (effective mass $m^{\star}=0.067 m_{e}$), as obtained by selective underetching of pseudomorphically strained bilayer [c.f. Fig.~\ref{fig:fig1}(a)] or multilayer semiconducting thin films \cite{Mendach-apl,Mendach-chapter,qhe1, gmr, qhe2}. We consider the length of the tube along its axis to be smaller than the characteristic mean free path of the high-mobility curved 2DEG ($l \simeq 10 \,\mu$m in (Al,Ga)As heterojunctions \cite{qhe1}). 
We also assume the inner radius of curvature, $R_{in}$, and the outer radius of curvature, $R_{out}=R_{in} + w \, d$ where $w$ is the winding number and $d$ the total thickness of the layer stack in which the 2DEG is embedded, to be $R_{in,out} \simeq 100$ nm, and thus comparable to the large Fermi wavelength $\lambda_F \simeq 40$ nm of the low-density electron gas. Such a compact 3D nanoarchitecture can also be viewed as a curved quantum point contact (QPC) \cite{qtransport,qpcprl} with circular cross section and width corresponding to the diameter of the nanotube. In this regime, quantum-size effects appear that lead to ballistic conductance quantization both at zero-field and in the presence of a magnetic field \cite{qpcprl}. 
In the latter case  the nature of the quantum states, and thus the ballistic conductance itself, strongly depends on whether one considers open or periodic boundary conditions.


\begin{figure}[tbp]
\includegraphics[width=.9\columnwidth]{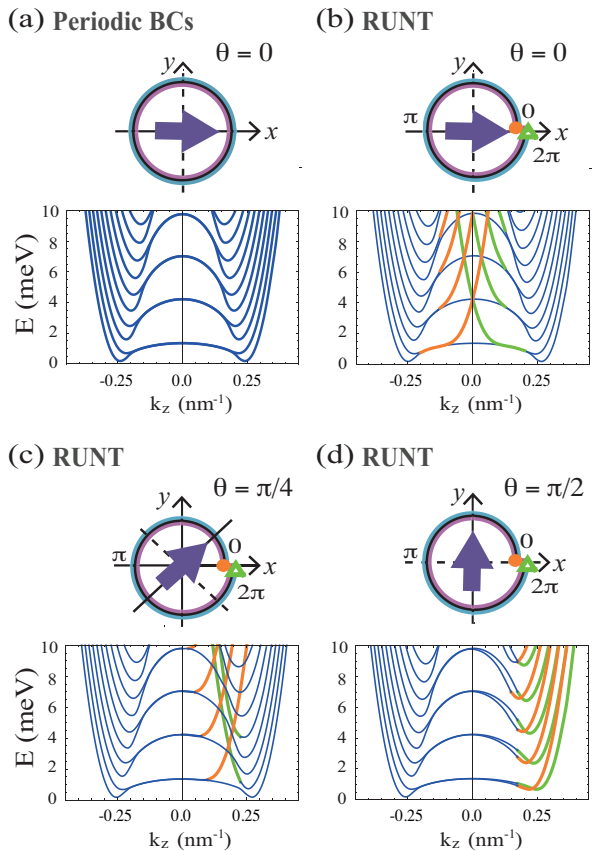}
\caption{(Color online). 
(a) Magnetic spectrum for a nanotube with $R_{in}=32 \pi\,$nm subject to a $1.65$ T transversal magnetic field  assuming periodic boundary conditions. $k_z$ is the momentum along the tube axis.
(b) Same for a single wound RUNT with open boundary conditions and magnetic field oriented along the edge axis ($\theta \equiv 0$). The green and orange lines explicitly show the dispersive edge states absent assuming periodic boundary conditions.
(c) and (d) show the evolution of the magnetic spectrum of panel (b) varying the magnetic field direction to $\theta=\pi/4$ and $\theta=\pi /2$ respectively.} 
\label{fig:fig2}
\end{figure}

To show this, we first neglect the difference between the inner and the outer radius of the nanotube ($R_{in} \equiv R_{out} \equiv R$), and thus consider a cylindrical 2DEG with periodic boundary conditions in an external magnetic field \cite{nt1,gaasnt1}, for which the radius of curvature is much larger than the Landau magnetic length $l_B = \sqrt{\hbar / (eB)}$. Since the radius of curvature is in the hundreds of nanometers scale, this regime can be reached already if magnetic fields on the scale of one Tesla are externally applied. 
This also implies that one can neglect the Zeeman splitting removing the spin degeneracy and take only into account the orbital effect of the magnetic field.  
The large separation between the Landau magnetic length and the radius of curvature allows for the formation of cyclotron orbits centered at the two positions where the tangential plane of the nanostructure is orthogonal to the magnetic field direction [see Fig.~\ref{fig:fig1}(b)]. The appearance of cyclotron orbits is reflected, in turn, in the formation of doubly-degenerate quasi-1D Landau-like states. This is demonstrated in Fig.~\ref{fig:fig2}(a), where we show the energy spectrum of a 2DEG confined to a cylindrical surface \cite{costa,strain} under the influence of a transversal magnetic field of strength $1.65$ T \cite{prlsch,aq3} [see Supplemental Material \cite{supple}]. 
At large values of the momentum $k_z$ along the translationally invariant tube axis direction, the flat quasi-1D Landau-like subbands, however, start to acquire a characteristic parabolic dispersion [c.f. Fig.~\ref{fig:fig2}(a)]. One can identify these dispersive states as snake states \cite{mul92} centered, due to Lorentz force, at the two points where the tangential plane of the nanostructure is parallel to the magnetic field direction [c.f. Fig.~\ref{fig:fig1}(b)]. 
As we show in the Supplemental Material, the appearance of snake electron trajectories results from the fact that in  the immediate vicinity of the orbit centers the spatially inhomogeneous normal component of the externally applied magnetic field switches sign.

Next, we explicitly take into account the difference between the inner and outer radius of curvature, and, following Ref.~\onlinecite{runt}, consider the 2DEG confined to a cylindrical surface whose cross section is approximated by an Archimedean spiral with polar equation $r \left(\phi \right) = R_{in} + d \phi / ( 2 \pi)$. We assume periodic boundary conditions along the tube axis $\hat{z}$, and open boundary conditions with $\phi \in \left(0 , 2 \pi w \right)$ along the azimuthal direction. 
The qualitative change in the nature of the quantum states for the open geometry is shown in Fig.~\ref{fig:fig1}(b) where we sketch the different quantum states appearing in a single wound ($w=1$) rolled-up nanotube (RUNT),  
subject to a magnetic field  whose direction $\theta$ with respect to the edge axis  [c.f. Fig.~\ref{fig:fig1}(a)]  is set to zero. We emphasize that generally the winding number will take a non-integer value $w > 1$ due to the unavoidable presence of overlapping fringes as found in experiments \cite{firstn} and computer simulations \cite{zan07}. This, however, does not qualitatively change the main features of the magnetic spectra, which are entirely set by the presence, and not the precise location, of the open boundaries.

The presence of the boundaries does not influence the formation of the snake states discussed above. However, due to  the hard walls, one of the two cyclotron orbits encountered in the closed geometry fractionalizes into two skipping orbits for which a clear dispersion along $k_z$ is expected. Fig.~\ref{fig:fig2}(b) shows the ensuing energy spectrum with  $d \simeq 12.5$ nm, $R_{in} \simeq 100$ nm and a magnetic field strength, as before, of 1.65 T [see Supplemental Material]. One can identify a single non-degenerate quasi-1D Landau-like state now coexisting for $k_z \simeq 0$ with two dispersive edge states localized at the inner and outer radius of the RUNT, which distinctly differentiate the magnetic spectra of rolled-up tubular nanostructures with respect to seamless tubes as carbon nanotubes.

Having established the qualitative difference between the magnetic spectra with periodic and open boundary conditions, we now discuss the interplay between the location of the hard-wall boundaries and the direction of the externally applied magnetic field. Since, as mentioned above, the open curved geometry of a spiral RUNT breaks the rotational symmetry of the embedding 3D space, the features of the magnetic spectra are drastically altered as the direction of the magnetic field changes. This is immediately manifested in Fig.~\ref{fig:fig2} where we 
show the evolution of the magnetic spectrum of Fig.~\ref{fig:fig2}(b) by applying two consecutive 45$^{\circ}$ tilts of the magnetic field while keeping its strength constant. As shown in Fig.~\ref{fig:fig2}(c) the first 45$^{\circ}$ rotation of the magnetic field direction restores a quasi-1D Landau state doublet for $k_z \simeq 0$ due to the insensitiveness of the small radii cyclotron orbits centered at $\phi=\pi/4$ to the hard-walls. The same holds true for the snake states at large values of $k_z > 0$ whose orbits are centered at $\phi= 7 \pi /4$. On top of this, we find the two edge states with skipping orbits at the inner and outer radius of the nanotube to appear for intermediate values of momentum $k_z >0$ [c.f. the green and orange lines in Fig.~\ref{fig:fig2}(c)]. Considering instead the magnetic field direction perpendicular to the edge axis leads to the magnetic spectrum shown in Fig.~\ref{fig:fig2}(d). It strongly resembles the spectrum for the periodic boundary conditions of Fig.~\ref{fig:fig2}(a) with the following caveat: due to the hard-walls 
the snake states at large momenta $k_z>0$ are substituted by edge states.
This, however, does not change qualitatively the magnetic spectrum apart from an asymmetry in the dispersion around $k_z=0$ due to the different nature of the magnetic states, {\it i.e.} snake orbits for $k_z<0$ and skipping orbits for $k_z>0$. 

\begin{figure}[tbp]
\includegraphics[width=\columnwidth]{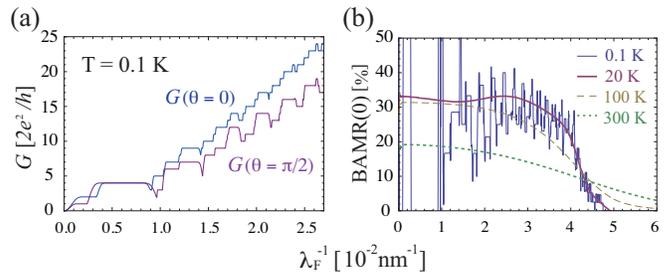}
\caption{(Color online). (a) Low-temperature magnetoconductance measured in $2 e^2 / h$ units as a function of the inverse of the 2DEG Fermi wavelength for a magnetic field oriented parallel to the edge axis (blue line) and perpendicular to it (purple line). (b) Maximum value of the anisotropy BAMR$(0)$ for temperatures up to room temperature.}
\label{fig:fig3}
\end{figure} 

The aforementioned features of the magnetic spectra persist up to a critical energy scale $E_c$ given by the characteristic Landau level energy splitting $\hbar \omega$, with $\omega$ the cyclotron frequency, renormalized by a geometrically tunable factor $R_{in}^2 / (2 \, l_B^2)$ [see Fig.~S2 in Supplemental Material]. This results from the following classical trajectories analysis: the existence of quasi-1D Landau levels is preserved as long as the radius of the cyclotron orbits $k_F l_B^2$ -- with $k_F$ the Fermi momentum  --  does not exceed the curvature radius. In the opposite regime, indeed, traversing trajectories \cite{qtransport} appear, and, independent of the magnetic field direction, edge and snake states coexist with magnetoelectric subbands. 
For a 2DEG in GaAs heterojunctions, we find the threshold between the two regimes for our chosen parameter set to be $\simeq 36$ meV -- an energy larger than $k_B T$ at room temperature.



With the magnetic spectra for different magnetic field directions in our hands, we have then determined the ballistic conductance of our single-wound nanotube neglecting intersubband scattering processes \cite{qpc-book, buttprl, static-C, Imry-book, ibm-qpc}. Within this approximation, the Landauer formula for the two-terminal conductance including the thermal smearing of the Fermi-Dirac distribution becomes \cite{japan-qc, bag-lg}
\begin{equation}
G(E_F, T,\theta)= \int_{0}^{\infty} G(E,0, \theta) \,\dfrac{\partial f}{\partial E_F} \, d E, 
\end{equation}
where $f$ indicates the Fermi-Dirac distribution, $E_F$ is the Fermi energy while $G(E,0,\theta) = 2 e^2 N(\theta) / h $ is the conductance at zero temperature proportional to the number $N (\theta)$ of occupied magnetoelectric subbands. Fig.~\ref{fig:fig3}(a) shows the behavior of the low-temperature magnetoconductance as a function of the inverse of the Fermi wavelength of the 2DEG $\lambda_F=\sqrt{2 \pi / n}$. Here $n$ is the electron density given by
\begin{equation}
n(E_F,T,\theta)=\frac{2}{L}\sum_i \int_{-\infty}^{\infty} \frac{dk_z}{2\pi}
f\left[E_i(k_z,\theta)-E_F,T \right],
\end{equation}
where the index $i$ runs over the occupied one-dimensional subbands, $L$ is the total arclength of the nanotube section and the factor of $2$ accounts for spin degeneracy. 
Independent of the direction of the externally applied magnetic field, the low-temperature magnetoconductance shows as a step-like increase, which is a direct consequence of the effective one-dimensionality of the curved nanoarchitecture. However, the different features of the magnetic spectra reported in Fig.~\ref{fig:fig2} lead to a substantial BAMR whose magnitude we define \cite{bal-ma} as
$$\text{BAMR}(\theta)=\dfrac{G (\theta) - G(\pi/2)}{G(\pi/2)}$$
measured from the reference direction perpendicular to the edge axis where the ballistic conductance takes its minimum value. 
In Fig.~\ref{fig:fig3}(b) we illustrate the relevance of the BAMR effect by showing that the magnitude at $\theta=0$ extends over a wide range of $\lambda_F$, and persists down to the critical Fermi wavelength $\lambda_F^{c} \simeq 2 \pi  l_B^2 / R$ ($ \simeq 25 \,$ nm for our set of parameters) where the cyclotron orbit's radius balances the radius of curvature.

\begin{figure}[tbp]
\includegraphics[width=\columnwidth]{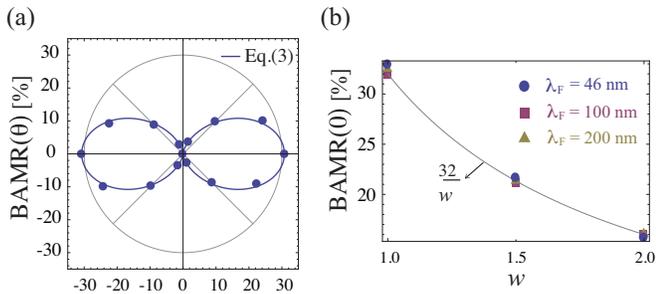}
\caption{(Color online). (a) Angular dependence of the BAMR at $T=20$ K  and Fermi wavelength $\lambda_F \simeq 46$ nm. The solid lines corresponds to a fit obtained using  Eq.~(\ref{eq:eqang}). (b) Scaling of the BAMR effect as a function of the number of windings of the nanotube for different values of the Fermi wavelength.}
\label{fig:fig4}
\end{figure}

The conductance steps disappear at temperatures $T > \hbar \omega / (4 k_B)$, {\it i.e.} when the width of the thermal smearing is larger than the subband splitting at the Fermi level. Remarkably, this thermal broadening does not endanger the occurrence of the BAMR. The onset of the anisotropy in the magnetoconductance is indeed regulated by the appearance of quasi-1D Landau states with cyclotron orbits, one of which, by tilting the magnetic field, fractionalizes into skipping orbits contributing to the conductance. As mentioned above, this occurs at the energy scale $E_{c}$, which is much larger than the subband splitting $\simeq 3$ meV. In Fig.~\ref{fig:fig3}(b) we prove indeed that a sizable BAMR survives at room temperature. Even more, as the effect of the thermal broadening increases, we find that the angular dependence of the BAMR [see Fig.~\ref{fig:fig4}(a)] can be accurately described by assuming the functional form of the resistivity for the classical anisotropic magnetoresistance effect \cite{bal-ma} 
\begin{equation}
\rho(\theta)=\rho(0) + \left[\rho(\pi/2) - \rho (0) \right] \sin^2{\theta}.
\label{eq:eqang}
\end{equation} 

On top of this, we find the BAMR effect to be
robust and {\it independent} of geometric details. We have indeed generalized our calculations to RUNTs with winding numbers $w=1.5,2$ [see Supplemental Material] and found that the BAMR magnitude $\propto 1/w$ [c.f. Fig.~\ref{fig:fig4}(b)], as it can be intuitively understood by considering that the number of snake and Landau states scales as $w$ while the number of edge states is {\it independent} of it.


To sum up, we have predicted the existence of a ballistic anisotropic magnetoresistance -- a change in the ballistic conductance with the direction of an externally applied magnetic field -- in compact rolled-up tubular nanostructures of conventional non-magnetic semiconducting materials \cite{first, runt-c, review-exp, si, si-sio}. The occurrence of this phenomenon is entirely due to  
the open curved geometry of these nanoarchitectures 
and is thus independent of geometric details of the nanostructures. 
A classical electron trajectories analysis shows that a  sizable BAMR occurs whenever the Fermi wavelength of the 2DEG $\lambda_F > 2 \pi l_B^2 / R$ and persists even in the classical regime up to a critical temperature 
$T_c \propto R^2 / l_B^4$.
As a result, a strong anisotropy in the magnetoconductance can be expected at room temperature and weak external magnetic fields in present day curved nanostructures manufactured with the rolled-up nanotechnology \cite{firstn,fet} where radii of curvatures in the hundreds of nanometer range can be achieved.

{\it Acknowledgements -- }
We acknowledge the financial support of the Future and Emerging Technologies (FET) programme within the Seventh Framework Programme for Research of the European Commission, under FET-Open grant number: 618083 (CNTQC). C.H.C. acknowledge support from the Ministry of Science and Technology of Taiwan (No: 102-2917-I-564-054).

\end{document}